      [1999/12/01 v1.4c Il Nuovo Cimento]
\documentclass{cimento}

\usepackage{graphicx}  
\usepackage{color}
  
\title{W+jets as a background to top physics: the quest for many jets}
\author{S.~Schumann\from{ins:x}}
\instlist{\inst{ins:x}{Institute for Theoretical Physics, University of Heidelberg\\ Philosophenweg 16, 69120 Heidelberg, Germany\\
email: {\tt s.schumann@thphys.uni-heidelberg.de}}} 
\PACSes{
\PACSit{}{\ 12.38.Bx,\ 12.38.Cy,\ 13.85.Hd,\ 13.87.-a}
}
\begin{document}

\maketitle

\begin{abstract}
The latest progress in calculating electroweak gauge boson production in 
association with QCD jets at hadron colliders is summarized. Particular 
emphasis is given to the recently completed QCD one-loop calculations 
of W+3jets and Wb final states. Furthermore recent developments in improving 
Monte Carlo event generators by means of combining tree-level matrix elements 
with parton showers is reviewed. 

\end{abstract}

\section{Introduction}

Top-physics is at the heart of the Tevatron and LHC physics programme. 
Since its first observation at the Tevatron in 1995 
\cite{Abe:1995hr,Abachi:1995iq} at lot of effort went into measuring 
top-quark production cross sections, its mass and quantum numbers. 
At the LHC top-quark physics can offer a unique window into potential 
new physics at the TeV scale, see for instance \cite{Frederix:2007gi}. 

\begin{figure}[th!]
\includegraphics[width=5.5cm,angle=-90]{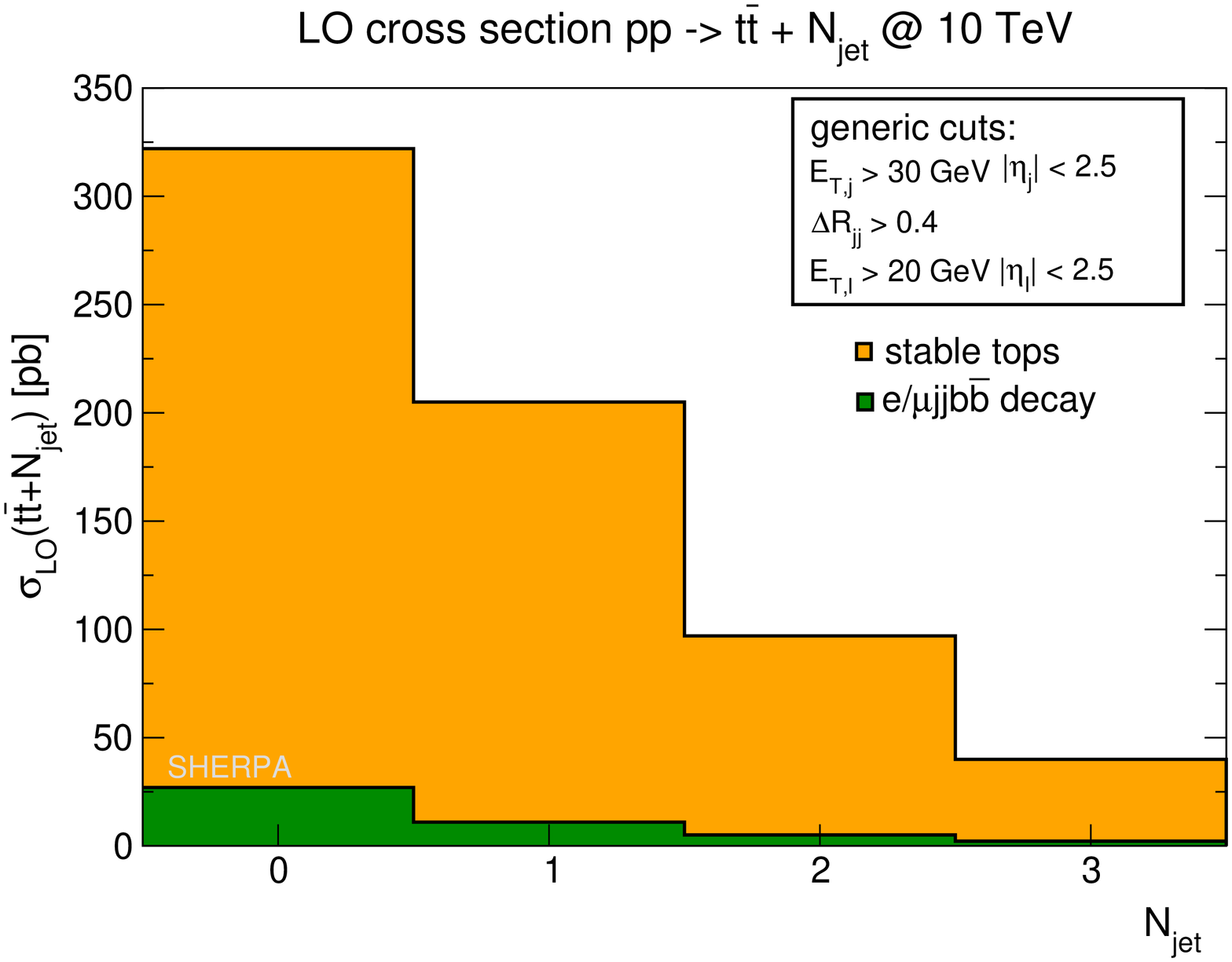}
\hfill
\includegraphics[width=5.5cm,angle=-90]{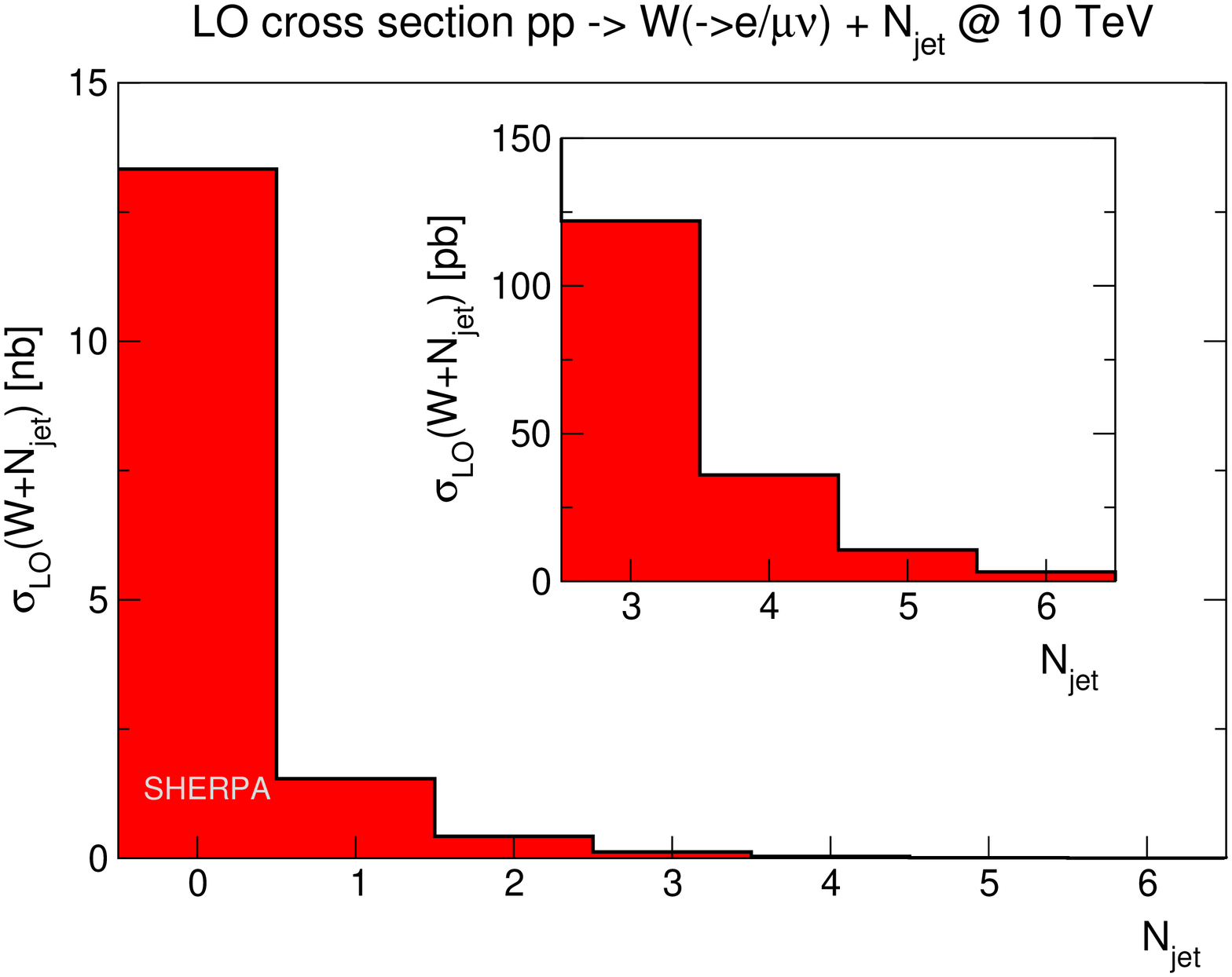}
\caption{The leading-order production cross sections for $t\bar t+n$-jets (left panel) and $W+n$-jets (right panel) in $pp$ collisions at $\sqrt{s}=10$ TeV. For the top signal the inclusive cross sections for stable top quarks and for the semi-leptonic decay channel are shown. For the latter and for the $W+$jets background a set of generic jet and lepton cuts have been applied. All cross sections have been calculated using {\sc Sherpa}~\cite{Gleisberg:2008ta} employing matrix elements from {\sc Comix} \cite{Gleisberg:2008fv}.\label{fig:tt_Wjets_xsecs}}
\end{figure}

When considering the semi-leptonic decay of a produced pair of 
top-quarks the resulting final state is $l^\pm+E_{T}^{miss} +$jets 
(where up to two jets might be heavy-flavor tagged). The very same 
signature is provided by a leptonically decaying $W$ associated by a 
corresponding number of QCD jets. To highlight the importance of the 
$W+n-$jets processes as the dominating background to top-pair 
production, the LHC production cross sections for $t\bar t+$jets 
and $W+$jets are presented in Fig. \ref{fig:tt_Wjets_xsecs}. 

Here only a generic set of cuts, namely $E_{T,j}>30$ GeV, 
$E_{T,l}>20$ GeV, $\Delta R_{j(j,l)} > 0.4$ and 
$|\eta_{j,l}|<2.5$, on the final-state leptons and jets (including 
those from the top decays) has been applied. From the left panel we 
can infer that the probability of producing a $t\bar t$ pair in association
with one or more extra jets is quite significant, an observation that is 
confirmed by the corresponding one-loop calculations 
\cite{Dittmaier:2008uj,Bevilacqua:2010ve}. These production
rates have to be confronted with the $W+n$-jets cross sections, displayed 
in the right panel. While the inclusive $W$ rate exceeds the $t\bar t$ cross section by orders of magnitude, when asking for $\geq 3$ jets the rates become of same size. However, even for $W+6$-jets, background to the semi-leptonic 
$t\bar t+2$-jets process, at leading order (LO) background exceeds the signal.

From these simple considerations it is evident that there is a strong 
demand for having predictions for the $W+n$-jets processes accurate at 
next-to-leading order (NLO) in QCD, reducing inherent scale uncertainties 
of the theoretical predictions. Furthermore an accurate modelling of this 
class of high jet-multiplicity processes in Monte Carlo event generators is
of major importance for the success of the ambitious LHC top-physics menu.
 
\section{W+3jets at next-to-leading order}

Until recently NLO predictions have been available only for final states 
involving a $W$ boson and up to two additional jets \cite{Campbell:2002tg}.
Significant progress in the evaluation of virtual matrix elements involving
many external legs has enabled two independent groups to eventually
calculate $W+3$jets at one-loop accuracy. 

In Refs. \cite{Ellis:2009zw,KeithEllis:2009bu} the leading-color 
approximation to the full result has been presented. In this calculation 
the D-dimensional generalized unitarity method as described in Ref. 
\cite{Giele:2008ve} is used to evaluate the loop amplitudes. The actual
calculation is performed in the framework of the {\sc MCFM} code \cite{mcfm}.
The authors of Refs. \cite{Ellis:2009zw,KeithEllis:2009bu} proposed 
a prescription called "leading color adjustment" that allows them to provide 
a sensible approximation to the full-color NLO result. In essence they 
rescale the leading color one-loop result by a constant factor defined to 
be the ratio of the LO full color cross section over its leading-color 
approximation. 

In Refs. \cite{Berger:2009zg,Berger:2009ep} the first complete NLO 
calculation of $W+3$jets has been presented. This calculation includes 
all partonic subprocesses and is exact in the treatment of color. For 
the one-loop matrix elements the program {\sc BlackHat} \cite{Berger:2008sj} 
is used that is based on unitarity methods 
\cite{Ossola:2006us,Forde:2007mi,Badger:2008cm}. For the generation of the
real-emission matrix elements, the Catani--Seymour dipole subtraction terms 
\cite{Catani:1996vz}, as well as all phase-space integrations the Monte 
Carlo generator {\sc Sherpa} \cite{Gleisberg:2008ta,Gleisberg:2007md} is used.

From NLO calculations we can expect a reduced dependence on the unphysical 
renormalization and factorization scales. However, they still exhibit a scale 
dependence.

\begin{table}[th!]
\begin{tabular}{|c||c|c|c|}
\hline \# of jets  & CDF & LO & NLO  \\
\hline 1 & $\; 53.5 \pm 5.6 \;$ & $ 41.40(0.02)^{+7.59}_{-5.94}$ &
           $\;  57.83(0.12)^{+4.36}_{-4.00} \;$ \\
\hline 2 & $ 6.8 \pm 1.1$  & $ 6.159(0.004)^{+2.41}_{-1.58}$ &
           $7.62(0.04)^{+0.62}_{-0.86} $  \\
\hline 3 & $ 0.84\pm 0.24$  & $ 0.796(0.001)^{+0.488}_{-0.276}$ & 
           \color{red}{$ 0.882(0.005)^{+0.057}_{-0.138}$}  \\
\hline
\end{tabular}
\caption{Total cross sections in pb with scale dependence for $W+n$-jets at Tevatron compared to data from CDF \cite{Aaltonen:2007ip}. Numbers taken from Ref. \cite{Berger:2009ep}, a complete description of the calculational setup and the cuts used can be found therein.\label{tab:Wjets_NLO_Tevatron}}
\end{table}

In Tab. \ref{tab:Wjets_NLO_Tevatron} the theoretical prediction for the $W+1,2,3$-jets cross 
sections calculated at LO and NLO are compared to a measurement by CDF \cite{Aaltonen:2007ip}. 
The inherent scale uncertainties are indeed significantly reduced for the one-loop results. The 
newly obtained $W+3$jets NLO result is in perfect agreement with the data. The predicted 
scale uncertainty for $W^\pm+3$jets production at the LHC is also largely reduced at NLO. 
Considering $pp$ collisions at 14 TeV and $E_T^{jet}> 30$ GeV, Ref. \cite{Berger:2009ep} quotes

\begin{eqnarray*}
\sigma^{\rm LO}_{W^-+3{\rm jets}}  = 22.28(0.04)^{+7.80}_{-5.34}\;{\rm pb} 
\quad&{\rm vs.}&\quad 
\sigma^{\rm NLO}_{W^-+3{\rm jets}} = 27.52(0.14)^{+1.34}_{-2.81}\;{\rm pb}\,,\\
\sigma^{\rm LO}_{W^++3{\rm jets}}  = 34.75(0.05)^{+12.06}_{-8.31}\;{\rm pb} 
\quad&{\rm vs.}&\quad
\sigma^{\rm NLO}_{W^++3{\rm jets}} = 41.47(0.27)^{+2.81}_{-3.50}\;{\rm pb}\,.
\end{eqnarray*}

\begin{figure}[bh!]
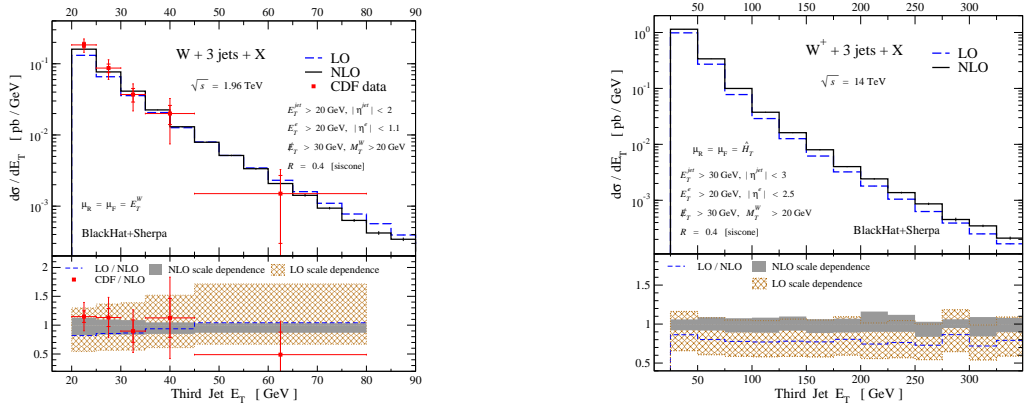

\includegraphics[width=5.5cm]{figs/W3jTev_ETWmu_siscone_eb_jets_jet_1_1_Et_3_with_CDF_data.eps}
\hfill
\includegraphics[width=5.5cm]{figs/Wp3jLHC_HTmu_siscone_eb_jets_jet_1_1_Et_3.eps}
\caption{Transverse-momentum distribution of the third-hardest jet in $W+3$jets events at the Tevatron, compared to data from CDF \cite{Aaltonen:2007ip} (left panel) and the LHC (right panel) at leading- and next-to-leading order. Figures taken from \cite{Berger:2009ep}.\label{fig:Jet3ET_NLO}}
\end{figure}

Besides a reduced scale dependence of the total $W+3$jets cross section the
NLO calculation exhibits largely narrowed uncertainty bands for differential 
distributions. This is exemplified in Fig. \ref{fig:Jet3ET_NLO}, where the 
transverse-momentum distribution of the third-hardest jet at Tevatron and LHC 
energies is shown. However, care has to be taken which central scale is 
actually used in this intrinsic multi-scale problem. As pointed out in 
Refs. \cite{KeithEllis:2009bu,Berger:2009ep} a choice like the bosons 
transverse momentum, $E_T^W$, can yield unphysical results for certain 
distributions, originating from large kinematic logarithms. A seemingly 
more appropriate choice 
is the total partonic transverse energy, $\hat{H}_T$.

\begin{figure}[th!]
\begin{center}
\includegraphics[width=5.5cm]{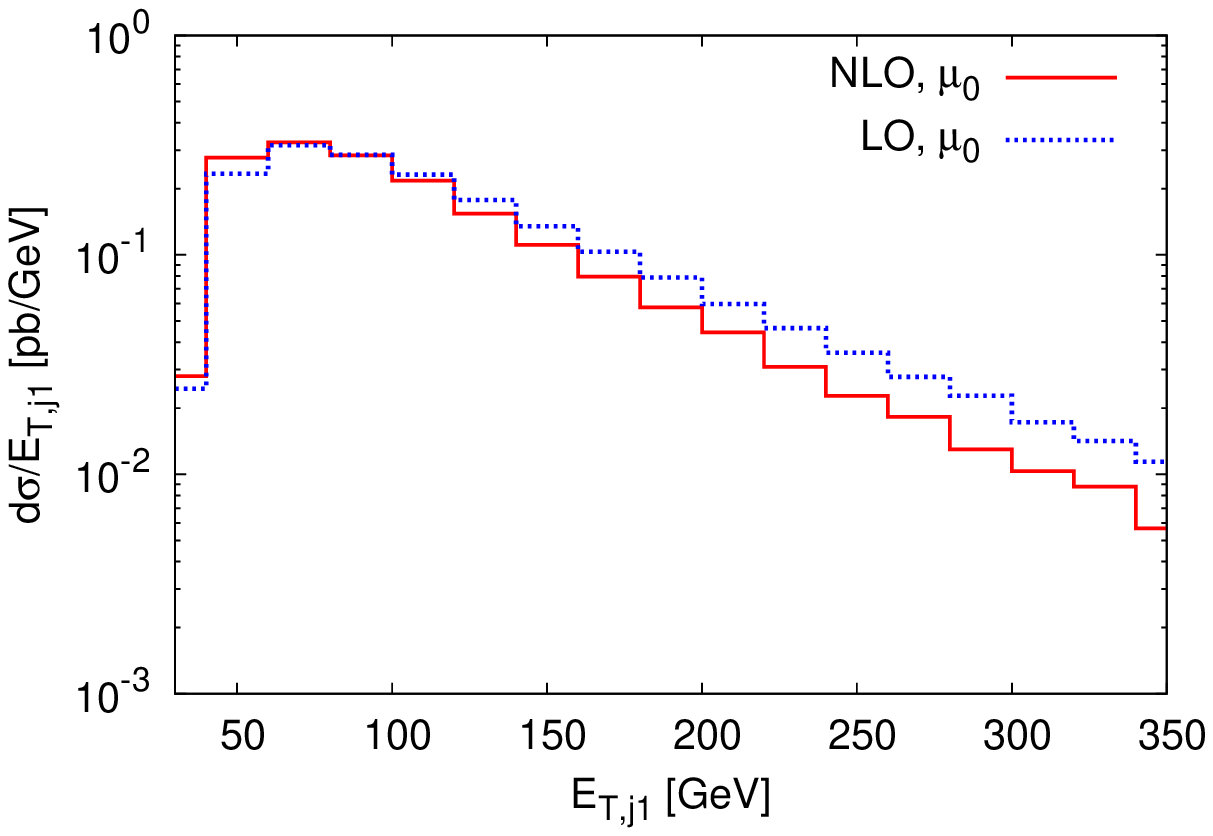}
\includegraphics[width=5.5cm]{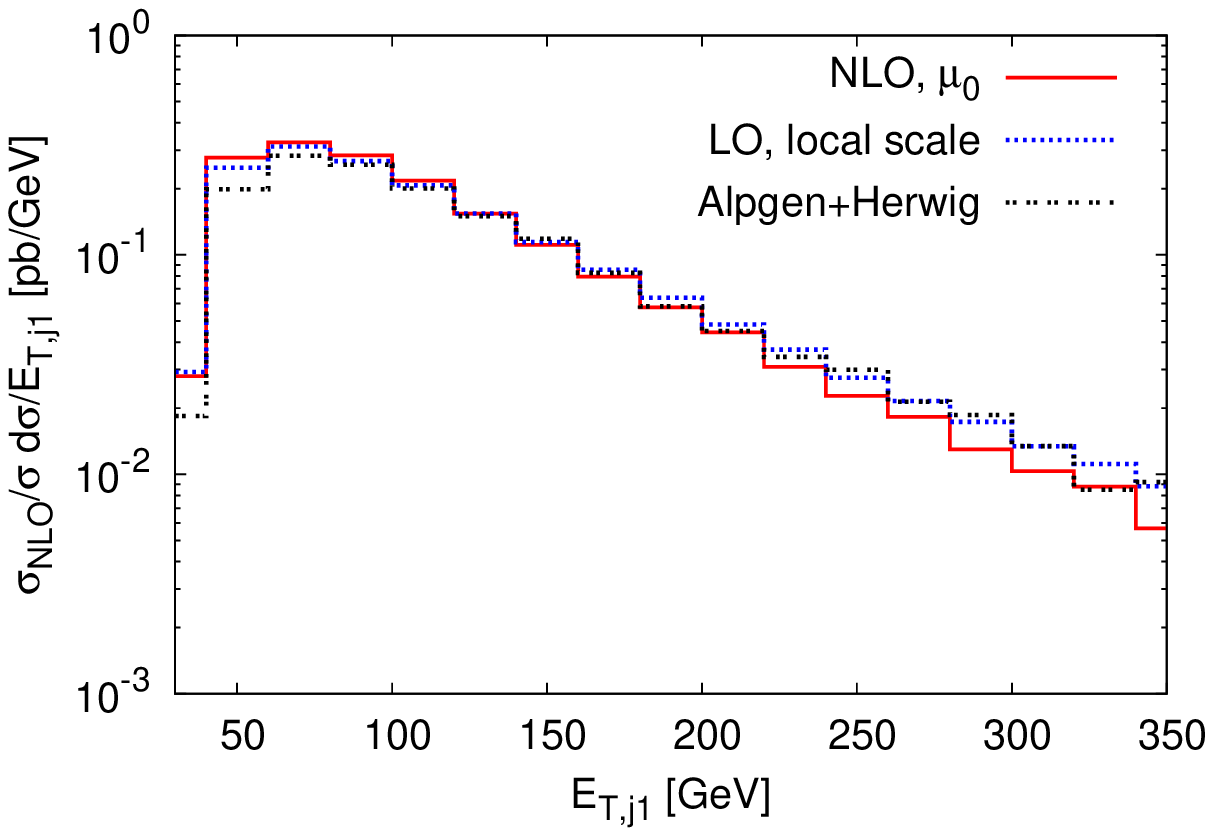}\\
\includegraphics[width=5.5cm]{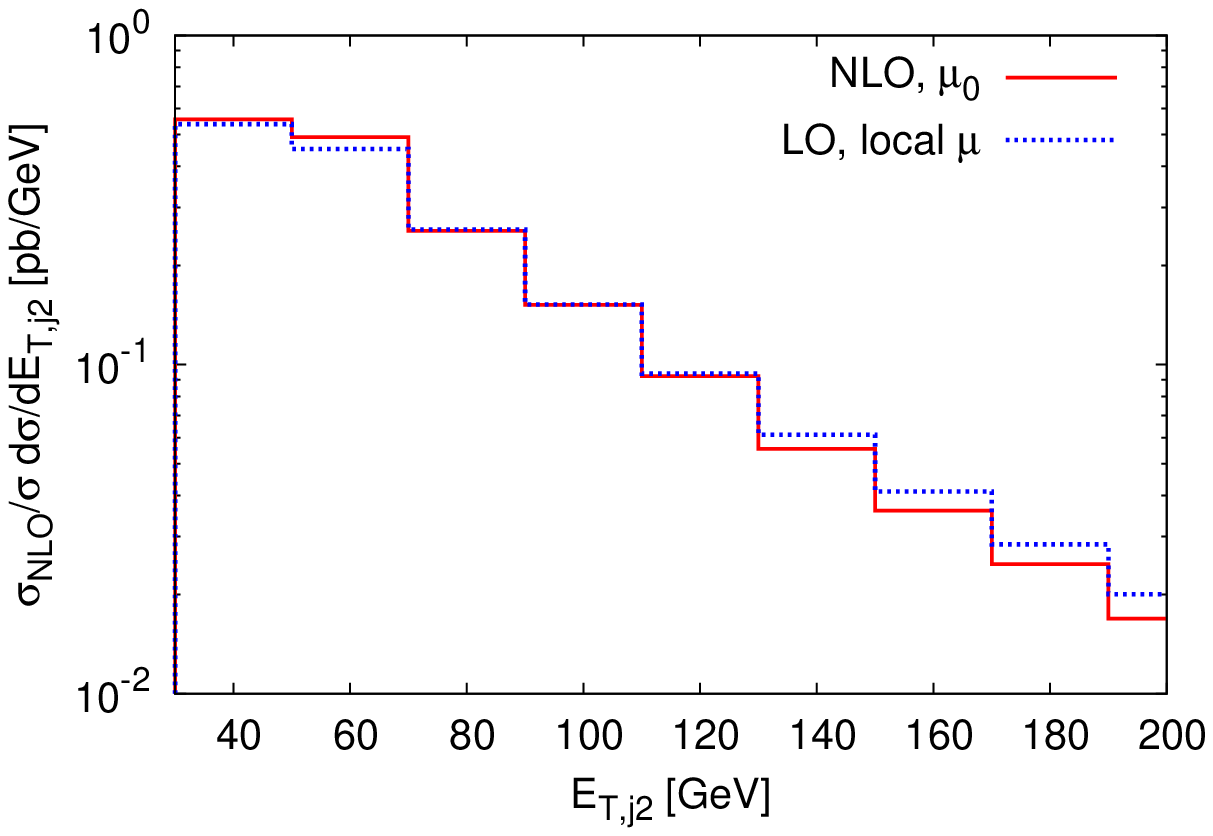}
\includegraphics[width=5.5cm]{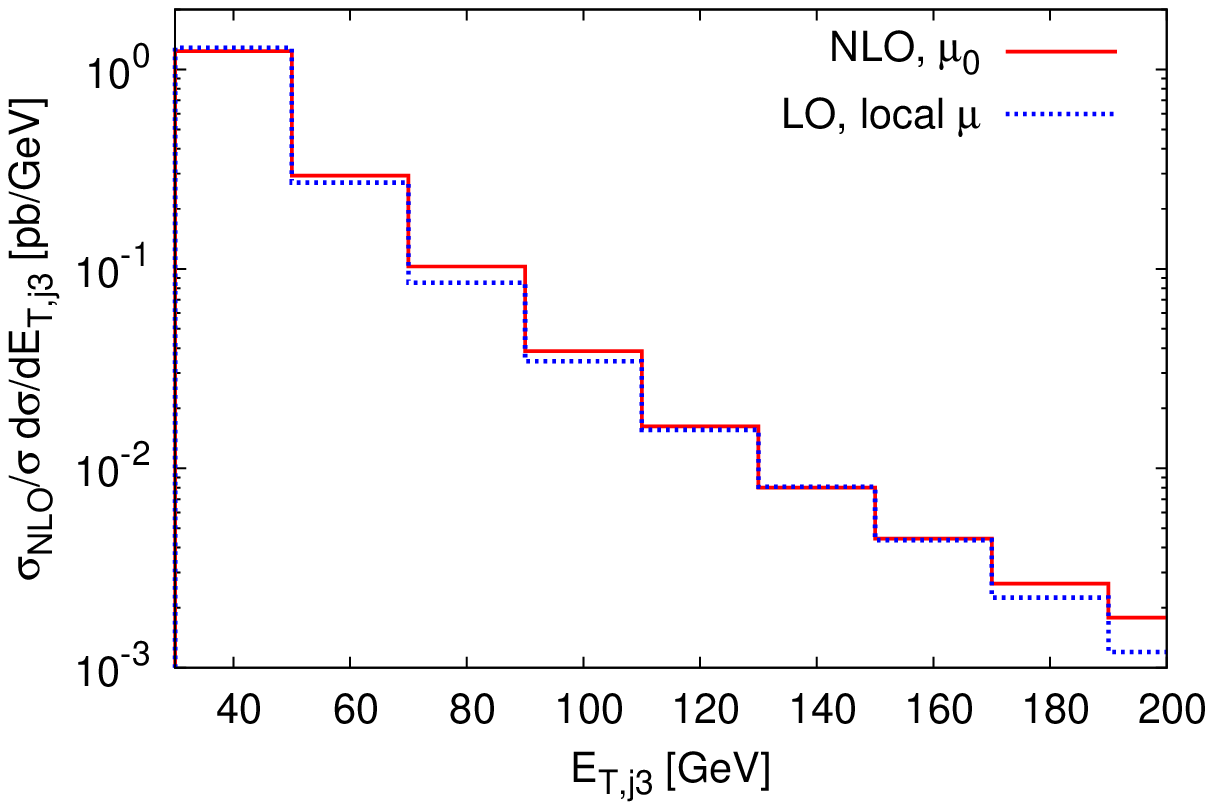}
\end{center}
\caption{Jet transverse-momentum distributions in $W^++3$jets production at a $10$ TeV LHC. The default scale is chosen to be $\mu_0=\mu_R=\mu_F = \sqrt{p^2_{T,W}+m^2_W}$. The local scale choice corresponds to taking each $\alpha_S$ factor in the LO calculation at its respective $k_T$ splitting scale. For details on the calculational setup see \cite{Melnikov:2009wh}.\label{fig:W3jets_localscales}}
\end{figure}

Ref. \cite{Melnikov:2009wh} discussed the possibility to accommodate shape 
differences between NLO and LO results by appropriately choosing 
scales for the strong coupling factors in the latter. In particular a 
local scale choice was investigated where each $\alpha_S$ factor is evaluated
at a reconstructed $k_T$ splitting scale. In Fig. \ref{fig:W3jets_localscales}
 a comparison between the default scale $\mu_0 = \sqrt{p^2_{T,W}+m^2_W}$ and the local prescription for the transverse-momentum distribution of 
the three hardest jets is shown. The local scale scheme is in much better 
agreement with the NLO shapes. This approach of local $\alpha_S$ factors 
is commonly used in parton shower Monte Carlos and in particular in 
calculations that combine tree-level multi-parton matrix elements with 
showers \cite{Catani:2001cc}. The explicit comparison for $W+3$jets final 
states at NLO confirms observations made in Refs. 
\cite{Krauss:2004bs,Krauss:2005nu} for $W+1,2$jets production and re-affirms 
the predictive power of the matrix element parton shower approach.
For a further study along these lines see \cite{SM_LH2009}.

\section{W+heavy flavors at next-to-leading order}

Concerning backgrounds to top-quark production special attention has to be 
given to $W+$jets final states with one or two jets being b-tagged. Using 
massive partons in the theoretical calculation removes corresponding soft 
and collinear singularities as they are regulated by the finite quark mass,
however at the price of the fixed-order calculation being more difficult. 
At present the NLO corrections for $Wb\bar b$, with massive $b$-quarks, 
are known \cite{Febres Cordero:2006sj}. When applying cuts that suppress 
contributions from the threshold region $m_{b\bar b}\approx 2m_b$ the actual
difference between the fully massive calculation and the limit $m_b=0$ is
typically less than 10\% \cite{Febres Cordero:2006sj}. 

\begin{figure}[th!]
\begin{center}
\includegraphics[width=5.cm,angle=90]{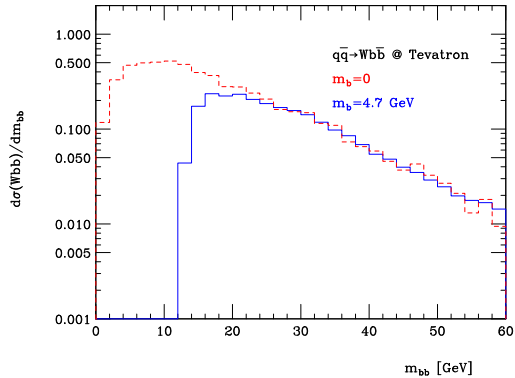}
\end{center}
\caption{LO differential cross section for $Wb$ production at Tevatron as a 
function of the $b\bar b$ invariant mass. Only one b-quark is found inside the fiducial volume here. Figure taken from \cite{Campbell:2008hh}, further details can be found therein. 
\label{fig:mbb_LO}}
\end{figure}

However, when sensitive to the threshold region or in case that just one 
heavy quark is tagged the massless approximation is not applicable. In the 
latter case the unobserved heavy jet must be integrated over the whole phase 
space thus introducing reference to the b-quark mass, 
cf. Fig. \ref{fig:mbb_LO}. One way out is to use heavy-quark parton 
distribution functions -- the so-called variable flavor scheme (VFS) that 
has the additional advantage to re-sum large logarithms of the type 
$\ln(m_W/m_b)$ to all orders. 

In Ref. \cite{Campbell:2008hh} a full NLO calculation of producing a $W$ boson
in association with just a single b-jet has been presented. This calculation 
consistently combines the massive $Wb\bar b$ calculation of 
\cite{Febres Cordero:2006sj} with the VFS computation of $Wbj$ \cite{Campbell:2006cu}. This calculation is an important ingredient when comparing the recent 
CDF measurement of the $W$ associated $b$-jet cross section \cite{Aaltonen:2009qi} with the NLO QCD calculation
\begin{eqnarray*}
\sigma^{\rm CDF}_{b-jets}\times {\cal{B}}(W\to l\nu) = 2.74^{+0.50}_{-0.50}\;{\rm pb} \quad{\rm vs.}\quad \sigma^{\rm NLO}_{b-jets}\times {\cal{B}}(W\to l\nu) = 1.22^{+0.14}_{-0.14}\;{\rm pb}\,.
\end{eqnarray*}
There is obviously tension between experiment and the theoretical results 
from NLO QCD as well as Monte Carlo predictions relying on matrix-element 
parton-shower merging \cite{Aaltonen:2009qi}. The source of this disagreement
is still under study - but might be assigned to the scale choice in the 
calculations \cite{Aaltonen:2008mt}. 

\section{Monte Carlo event generators}
In cases we lack a full NLO calculation (e.g. $W+\geq 4$jets) or observables 
are sensitive to multiple-parton emission and hadronization effects, 
theoretical predictions rely on the ability of multi-purpose Monte Carlo 
generators such as {\sc Pythia} \cite{Sjostrand:2006za}, {\sc Herwig} 
\cite{Bahr:2008pv} or {\sc Sherpa} \cite{Gleisberg:2008ta,Gleisberg:2003xi} 
to account for the underlying physics. Over the past decade enormous efforts 
went into improving these calculations by consistently incorporating multi-leg 
tree-level matrix elements into parton-shower simulations in the spirit of
\cite{Catani:2001cc,Lonnblad:2001iq,Caravaglios:1998yr}. For an overview 
of available approaches and an extensive comparison for $W+$jets production 
at Tevatron and LHC see Ref. \cite{Alwall:2007fs}. Essentially two major 
problems have to be addressed by each tree-level merging algorithm:
\begin{itemize}
\item How to attach a parton shower to a multi-leg tree-level matrix-element 
calculation without spoiling the logarithmic accuracy of the underlying QCD 
resummation?
\item How to avoid potential double- or under counting of phase-space 
configurations present in the parton shower and corresponding matrix-element 
calculations?
\end{itemize}

To accommodate these conditions in a generic tree-level merging algorithm
\begin{itemize}
\item multi-parton matrix elements get regularized through a suitably defined 
jet measure (e.g. a critical $k_T$- or cone-like distance);
\item appropriate starting conditions for the initial- and final-state 
parton shower have to be determined and certain (hard) shower emissions
need to be vetoed.
\end{itemize}

In particular the second item is subject to certain approximations in the
various schemes. An important concept to overcome those approximations is
a so-called {\em truncated shower}, first proposed in Ref. 
\cite{Nason:2004rx}. The underlying observation is that due to a mismatch 
of the jet-measure, used to slice the emission phase space, and the actual 
shower-evolution variable the radiation pattern of soft/large-angle 
emissions can be distorted. 

\begin{figure}[b!]
\begin{center}
\includegraphics[width=5.5cm]{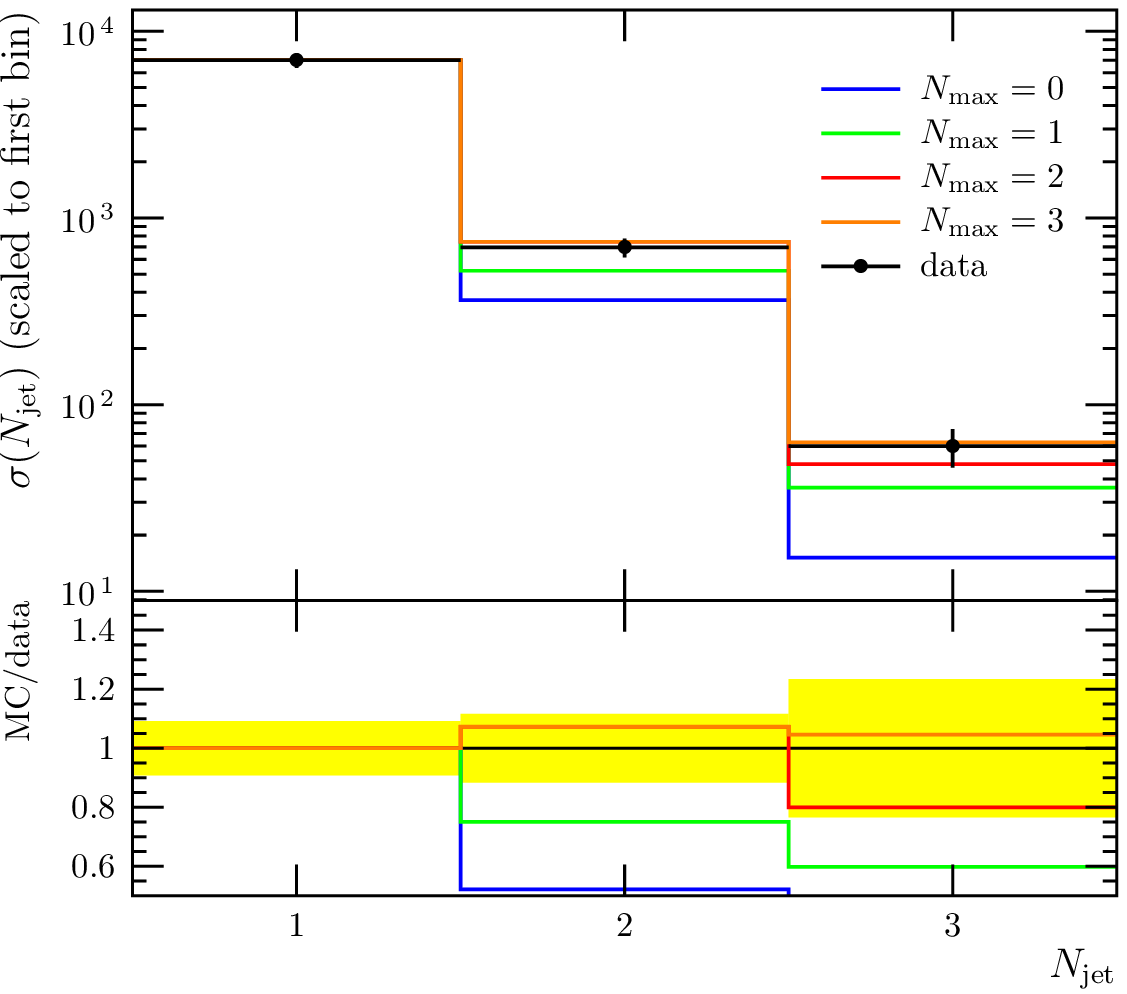}
\hfill
\includegraphics[width=5.5cm]{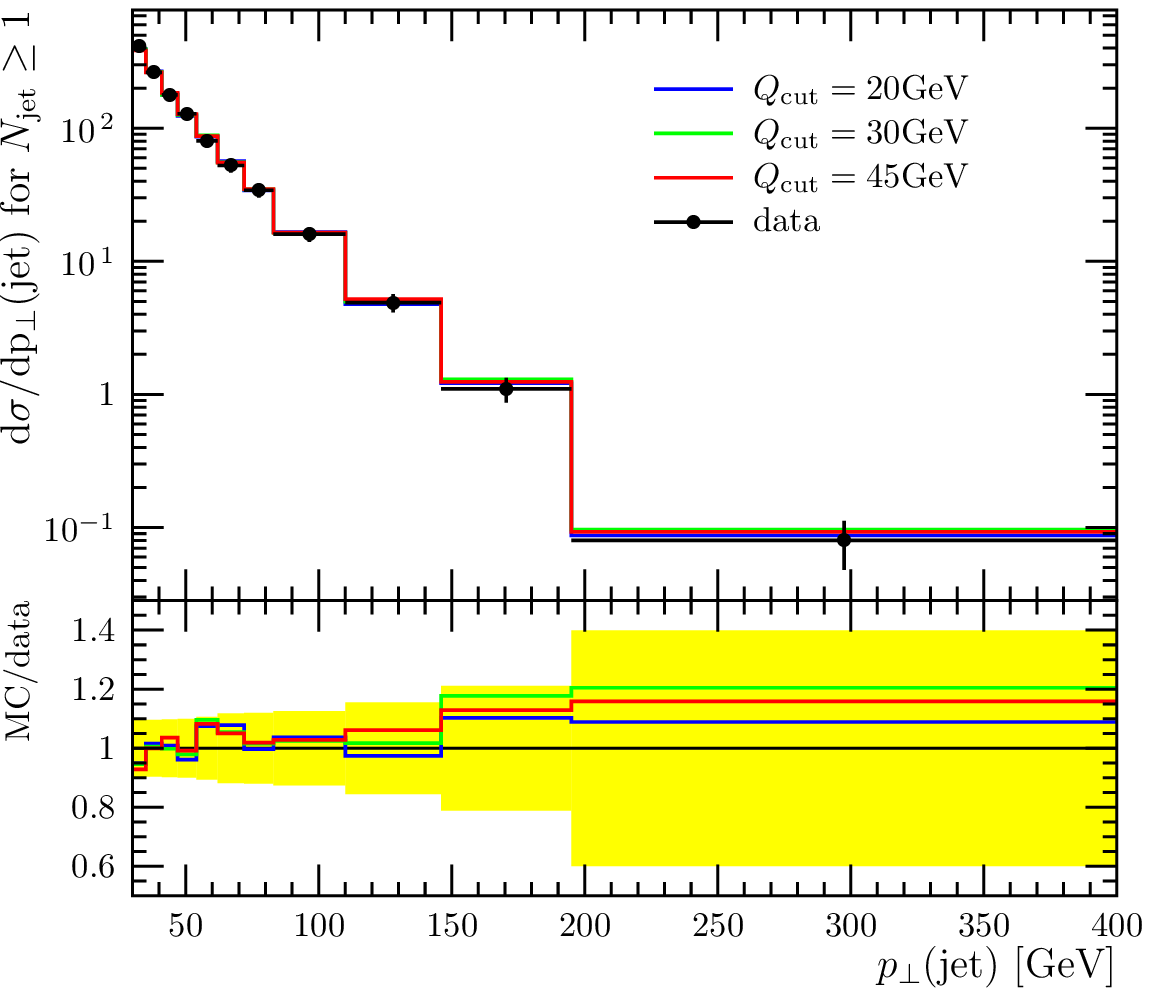}
\end{center}
\caption{Jet multiplicity (left panel) and the leading jet $p_T$ 
spectrum (right panel) in inclusive $Z^0/\gamma^*+$jets production 
compared to data from CDF \cite{Aaltonen:2007cp}. Figures taken from 
Ref. \cite{Hoeche:2009rj}\label{fig:METS_vs_data}}
\end{figure}

\begin{figure}[t!]
\begin{center}
\includegraphics[width=5.5cm]{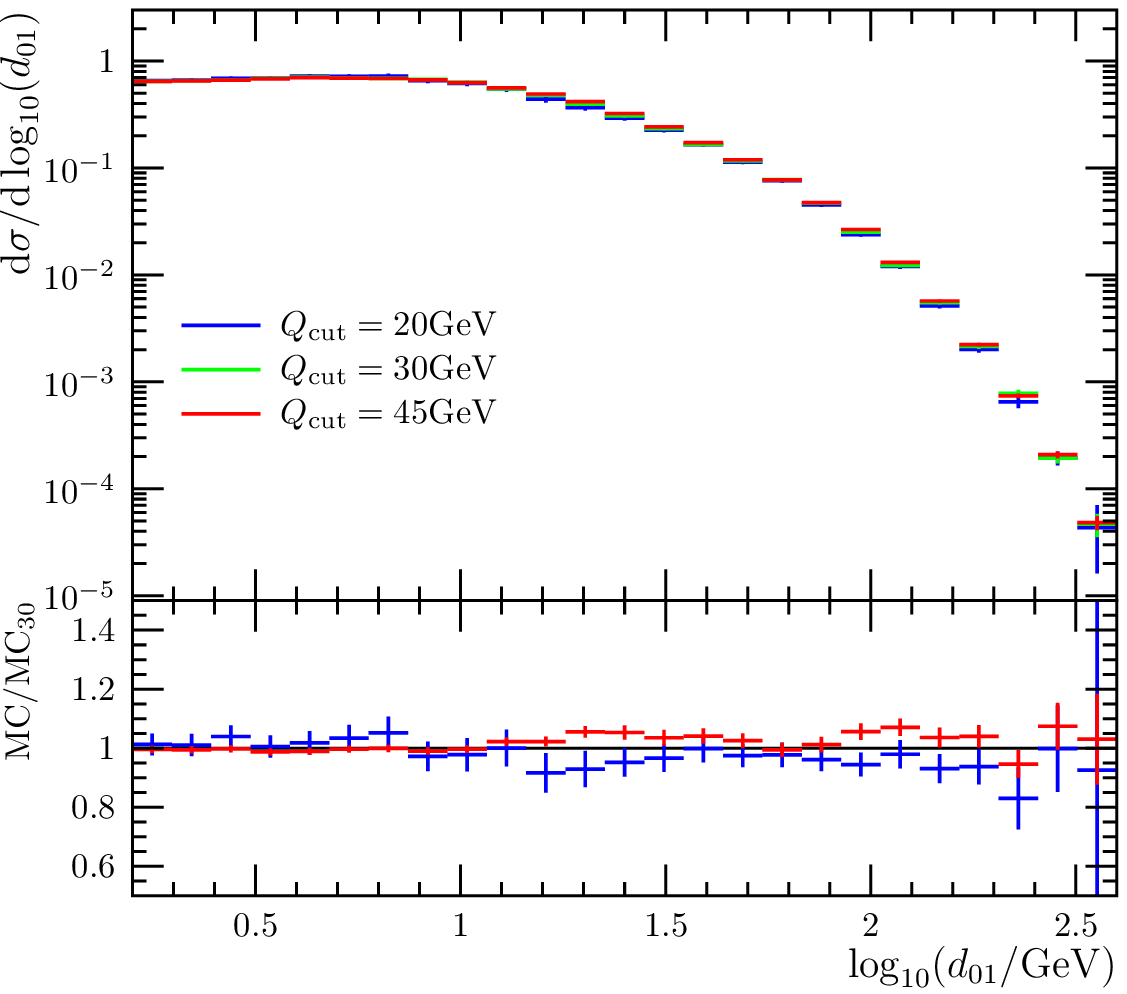}
\hfill
\includegraphics[width=5.5cm]{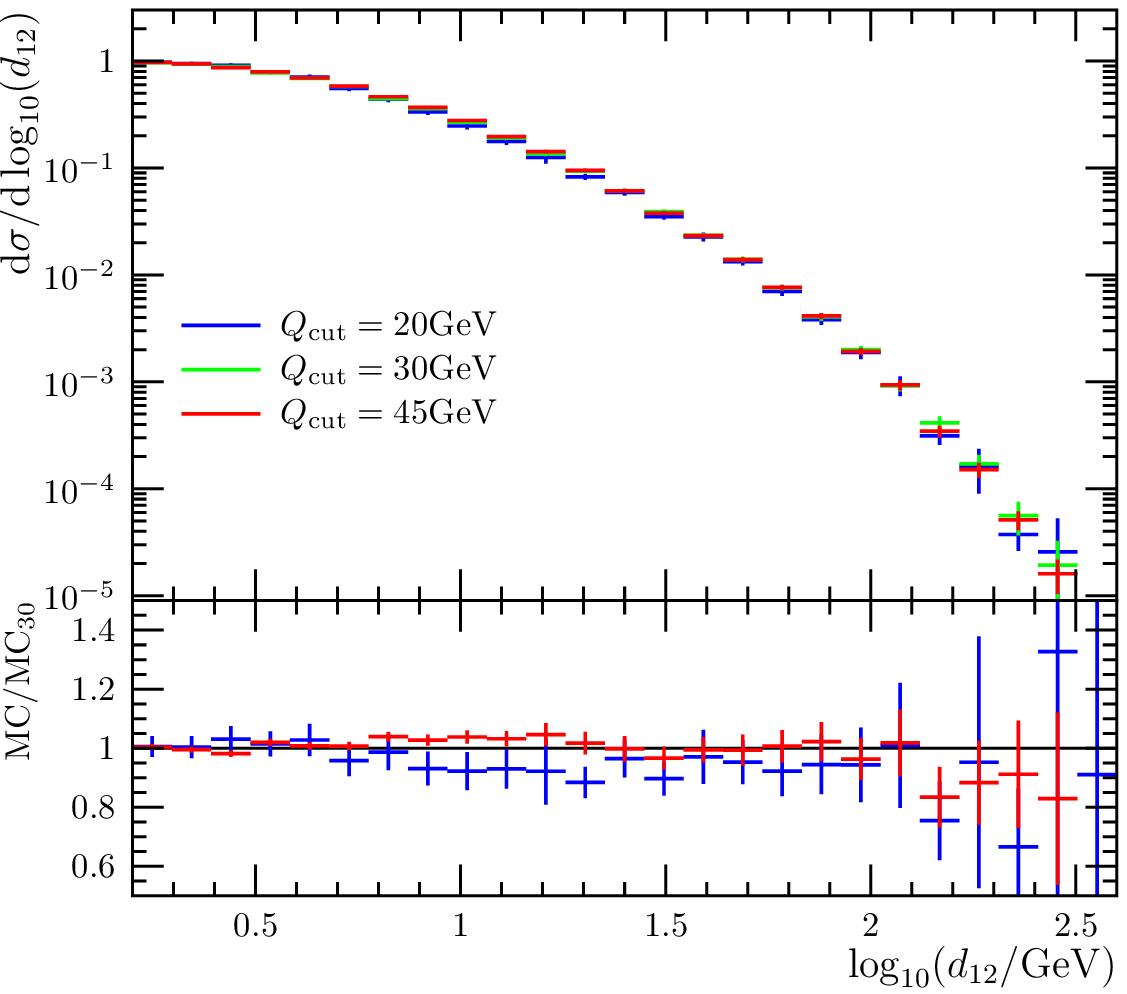}
\end{center}
\caption{Differential jet rates $d_{01}$ (left panel) and $d_{12}$ 
(right panel) for the CDF Run II $k_T$-algorithm \cite{Blazey:2000qt}. Displayed are the 
predictions for three different values of the merging cut. Figures taken 
from Ref. \cite{Hoeche:2009rj}\label{fig:METS_systematics}}
\end{figure}

In Ref. \cite{Hoeche:2009rj} such a truncated 
shower was implemented for the first time. The implementation relies on the
shower algorithm based on Catani--Seymour dipole factorization 
\cite{Schumann:2007mg} and combines it with the matrix-element generators
available inside the {\sc Sherpa} framework. The method has successfully been 
applied to jet production in $e^+e^-$ collisions, the Drell-Yan process 
\cite{Hoeche:2009rj}, prompt-photon production \cite{Hoeche:2009xc} and 
deep-inelastic scattering \cite{Carli:2010cg}. As of version 1.2 it 
constitutes the default method for combining matrix elements with parton 
showers in the {\sc Sherpa} generator. The new merging approach yields a 
largely reduced dependence on the intrinsic merging parameters compared to 
the previous CKKW implementation in {\sc Sherpa} and other merging algorithms \cite{Alwall:2007fs}. This is illustrated by the systematics studies for $Z^0/\gamma^*+$jets production at Tevatron presented in Figs. \ref{fig:METS_vs_data} and \ref{fig:METS_systematics}. The first figure presents a comparison of the
jet multiplicity and leading-jet $p_T$ distribution while in the latter 
the variation of the $k_T$ differential jet rates $d_{1\to 0}$ and $d_{2\to 1}$ for three different values of the slicing measure are presented. 

\section{Conclusions and Outlook}
Understanding the process of electroweak gauge boson production in association
with QCD jets is crucial for the success of the top-physics programme
both at the Tevatron and even more so at the LHC. In the last few years 
there has been enormous progress in the calculation of one-loop corrections 
to multi-parton final states. As a result the processes $W+3$jets and 
$Z^0/\gamma^*+3$jets are meanwhile known at next-to-leading order in QCD. 
In fact Ref. \cite{Berger:2010gf} already reports on first steps towards 
the calculation of $W+4$jets at the one-loop level using the 
{\sc BlackHat}+{\sc Sherpa} package. At this conference M. Worek reported 
on the {\sc  Helac-Nlo} package, that has proven to be capable of doing 
calculations of this complexity as well and next-to-leading order calculations 
for $Wb\bar b+\leq3$jets now seem to be feasible.

Concerning the simulation of $W+$jets with Monte Carlo event generators a 
high level of sophistication has been reached. The approach of combining 
multi-leg tree level matrix elements with parton showers has become a widely 
used standard that delivers results in good agreement with data from Tevatron
and exact higher-order calculations. One important future direction will be
to precisely understand how these methods can be generalized to allow for the
inclusion of one-loop matrix elements. First proposals in this direction 
have been made and implemented already, cf. Refs. 
\cite{Lavesson:2008ah,Hamilton:2010wh} and P. Nason's contribution to these 
proceedings. A novel procedure how to combine next-to-leading calculations 
of different final-state multiplicity, though not facing the problem of 
attaching parton showers, has been presented in Ref. \cite{Rubin:2010xp}.

\acknowledgments
The author would like to thank the organizers for the invitation and in 
particular Fabio Maltoni. Financial support from BMBF is gratefully acknowledged. 

\end{document}